\documentclass[epsfig,indentfirst,bm,12pt]{article}
\usepackage{CJK}
\usepackage{graphicx,amsmath,dcolumn,bm}
\usepackage{color}
\begin{document}
\title{\Large{\bf{The energy loss effect of incoming gluon from $J/\psi$ production in p-A collisions }}
\thanks {Supported partially by National Natural Science Foundation of China (11405043, 11575052).}}

\begin{CJK*}{GBK}{song}
\author{Li-Hua Song $^{1}$
\footnote{\tt{ E-mail:songlh@ncst.edu.cn}}
lin-Wan Yan $^{1}$
Chun-Gui Duan $^{2}$}

\date{}

\maketitle

\end{CJK*}

\noindent {\small 1.College of
Science,  North China University of Science and Technology, Tangshan 063009,
China}\\
{\small 2.Department of Physics, Hebei Normal
             University,
             Shijiazhuang 050024, China}

\baselineskip 9mm
\begin{abstract}
The energy loss effect of incoming gluon from $J/\psi$ production in p-A (or d-A) collisions is investigated by means of the E866, RHIC and LHC experimental data. The gluon mean energy loss per unit path length $dE/dL = 2.18\pm0.14$ GeV/fm is extracted by fitting the E866 experimental data for $J/\psi$ production cross section ratios $R_{W(Fe)/Be}(x_{F})$.
The obtained result indicates that the incoming gluons lose more energy than the incident quarks.
 By comparing the theoretical results with E866, RHIC, and LHC experimental data, it is found that the nuclear suppression due to the incident gluon (quark) energy loss reduces (increases) with the increase of the kinematic variable $x_{F}$ (or $y$). The energy loss effect of incoming gluon plays an important role on the suppression of $J/\psi$ production in a wide energy range from $\sqrt{s}=38.7$ GeV to $\sqrt{s}=5.0$ TeV, and the influence of incident quark energy loss can be ignored for high energy(such as at RHIC and LHC energy).

\vskip 1.0cm

\noindent{\bf Keywords:} $J/\psi$ production, gluon, charm quark, energy
loss.

\noindent{\bf PACS:} 24.85.+p; 
                     25.40.-h;  
                     12.38.-t; 
                     13.85.-t; 

\end{abstract}

\maketitle
\newpage
\vskip 0.5cm

\section{Introduction}

In order to quantify the properties of the QGP created in heavy-ion collisions, a solid understanding of the
nuclear modification of particle spectra in cold nuclear matter is fundamentally important. $J/\psi$ production in
proton-nucleus collisions provide an ideal tool to test the microscopic dynamics of medium-induced parton energy loss.

Drastic nuclear suppression effects are observed in a wide collision energy range for minimum bias p-A and d-A collisions, such as NA3[1], E772[2],
E866[3,4], NA50[5], HEAR-B[6], LHC[7,8] and RHIC[9]experiments. However, it is striking that there is no consensus on the origin of $J/\psi$ suppression in some kinematical conditions[10]. Some approaches attribute $J/\psi$ suppression to an effective absorption cross section $\sigma_{abs}$ of the $c\overline{c}$ pair[11-12]; other models attribute $J/\psi$ suppression to the increase of the $c\overline{c}$ pair mvariant mass by the multiple soft rescatterings through the nucleus, leading to a reduction of the overlap with the $J/\psi$ wave function[13].

In the nucleus rest frame, a high-energy $J/\psi$ is formed long after the nucleus thus what actually propagates through the nucleus is the parent $c\overline{c}$ pair. Our previous works[14-15] support that the nuclear modification on the parton distribution functions and the incident proton energy loss owing to multiple scattering on the surrounding nucleon and gluon radiation are the main initial state effects induced the $J/\psi$ suppression, and the energy loss of color octet $c\overline{c}$
is the dominant final state effect when the $c\overline{c}$ pair
remains colored on its entire path in the medium. In paper[15], by using
the EPS09 nuclear parton distributions [16] together with the energy loss of
the proton beam in initial state (The center-of-mass system
energy loss per collision $\triangle\sqrt{s}=0.18$ GeV
is determined from the nuclear Drell-Yan experimental data in the
Glauber model[17].) and the linear quark energy
loss in final state, we extracted the charm quark mean energy loss per unit path length ($dE/dL= 1.49\pm0.37$GeV/fm with $\chi^{2}/ndf=0.91$) by fitting the E866 experimental data\textcolor[rgb]{1.00,0.00,0.00}{[4]} in the region $0.2 <x_F < 0.65$.

To further investigate the microscopic dynamics of medium-induced parton energy loss, the color charge of parton energy loss has received significant interest. This issue is of fundamental importance for accurately understanding the dynamics for modifying a hard probe and the dense QCD properties of what is probed. Previous research [18-22] predict that: gluons lose more energy than quarks because of the stronger coupling to the medium.

In the $J/\psi$ production for p-A (or d-A) collisions, the observed suppression induced by the incident parton energy loss effect can give a better way to discriminatingly identify the energy loss of incoming gluon and quark. Following our previous work, in the present study we investigate the incoming gluon energy loss effect by means of the E866\textcolor[rgb]{1.00,0.00,0.00}{[4]}, RHIC[9] and LHC[7,8] experimental data, and desire that our research
can provide useful reference for deep understanding the microscopic dynamics of medium-induced parton energy loss.

The remainder of this paper is organized as follows. In Section II, the theoretical framework of our study is introduced. Section III is devoted to the results and discussion. Finally, a summary is presented.

\section{The formalism for $J/\psi$ production differential cross sections }
In the the color evaporation model (CEM)[23], for $J/\psi$
production in p-A collisions quarkonium production is
treated identically to open heavy-quark production except that the
invariant mass of the heavy quark pair is restricted to be less than
twice the mass of the lightest meson that can be formed with one
heavy constituent quark. For charmonium the upper limit on the
$c\overline{c}$ pair mass is then $2m_{D}$. The hadroproduction of heavy quark at leading order
(LO) in perturbative QCD is the sum of contributions from
$q\overline{q}$ annihilation and $gg$ fusion. The charmonium
production cross section ${d\sigma}_{p-p}/{dx_{F}}$ is a convolution
of the $q\overline{q}$ and $gg$ partonic cross sections with the parton distribution
functions $f_{i}$ in the incident proton and $f'_{i}$ in the target proton and expressed as [24]:
\begin{eqnarray}
\frac{{d\sigma}_{p-p}}{{dx_{F}}}(x_{F})&=&\rho_{J/\psi}\int^{2m_{D}}_{2m_{c}}dm\frac{2m}{\sqrt{x_{F}^{2}s+4m^{2}}}
\times[f_{g}(x_{1},m^{2})f'_{g}(x_{2},m^{2})\sigma_{gg}(m_{c}^{2})\nonumber \\
& & +\sum^{}_{q=u,d,s}\{f_{q}(x_{1},m^{2})f'_{\bar{q}}(x_{2},m^{2})
+{f_{\bar{q}}(x_{1},m^{2})f'_{q}(x_{2},m^{2})}
\}\sigma_{q\bar{q}}(m_{c}^{2})].
\end{eqnarray}
Here, in the rest frame of the target nuclei, $x_{1(2)}$ is the projectile proton (target) parton momentum
fractions, $x_{F}=x_{1}-x_{2}$, $\sqrt{s}$ is the center of mass
energy of the hadronic collision, $m^{2}=x_{1}x_{2}s$, $m_{c}=1.2$
GeV and $m_{D}=1.87$ GeV are respectively the charm quark and D
meson mass, $\sigma_{gg}(\sigma_{q\bar{q}})$ is the LO
$c\overline{c}$ partonic production cross section from the gluon
fusion (quark-antiquark annihilation), and $\rho_{J/\psi}$ is the
fraction of $c\overline{c}$ pair which produces the $J/\psi$ state.

In $J/\psi$ production from p-A (or d-A) collisions, owing to multiple scattering on the surrounding nucleon and gluon radiation while incident parton propagating
through the nucleus, the incoming gluon (quark) can lose its energy $\Delta E_{g}$ ($\Delta E_{q}$). The energy loss of incoming gluon (quark) results in an average change in its momentum fraction prior to the collision,
\begin{equation}
\Delta x_{1g}=\Delta E_{g}/E_{p},
\Delta x_{1q}=\Delta E_{q}/E_{p}.
\end{equation}
According to the parametrization for parton energy loss [25-26], the mean energy loss of incoming gluon (quark) can be expressed as:
\begin{equation}
\Delta E_{g}=\alpha L_{A}, \Delta E_{q}=\beta L_{A}.
\end{equation}
Here,  $L_{A}=3R_{A}/4 $ $(R_{A}=1.12A^{1/3})$ [27], and $\alpha$, $\beta$ are the parameters that can be extracted from
the experimental data by adopting the $\chi^2$ analysis method.

When the $J/\psi$ hadronization occurs
outside the nucleus, nuclear absorption should play little or no role and the energy loss of color octet $c\overline{c}$ is the dominant final state effect. In view of the shift in $x_{F}$ due to the energy loss of color octet $c\overline{c}$ ($\Delta E_{c\bar{c}}$), the momentum fraction of the incident gluon (quark) is actually:
\begin{equation}
x'_{1g}=x'_{1}+\Delta x_{1g}, x'_{1q}=x'_{1}+\Delta x_{1q},
\end{equation}
with $x'_{1}=\frac{1}{2}[\sqrt{x'^{2}_{F}(1-\tau)^{2}+4\tau}+x'_{F}(1-\tau)]$, $x'_{F}=x_{F}+\Delta E_{c\bar{c}}/E_{p}$[15], $\tau=m^{2}/s$.
The $J/\psi$ differential production cross section in p-A collisions ${d\sigma}_{p-A}/{dx_{F}}$ is written as:
\begin{eqnarray}
\frac{{d\sigma}_{p-A}}{{dx_{F}}}(x_{F})&=&\rho_{J/\psi}\int^{2m_{D}}_{2m_{c}}dm\frac{2m}{\sqrt{x_{F}^{2}s+4m^{2}}}
\times[f_{g}(x'_{1g},m^{2})f'_{g}(x'_{2},m^{2})\sigma_{gg}(m^{2})\nonumber \\
 & &+\sum^{}_{q=u,d,s}\{f_{q}(x'_{1q},m^{2})f'_{\bar{q}}(x'_{2},m^{2})
+{f_{\bar{q}}(x'_{1q},m^{2})f'_{q}(x'_{2},m^{2})}
\}\sigma_{q\bar{q}}(m^{2})].
\end{eqnarray}
Here, in consideration of the shift in $x_{F}$ due to the energy loss of color octet $c\overline{c}$, the target parton momentum fraction is actually
$x'_{2}=\frac{1}{2}[\sqrt{x'^{2}_{F}(1-\tau)^{2}+4\tau}-x'_{F}(1-\tau)]$.

Further, considering the energy loss of incident gluon, incoming quark and the color octet $c\overline{c}$, the leading order for $J/\psi$ production cross section as a function of $y$ should be written as:
\begin{equation}
\frac{{d\sigma}_{p-A}}{{dy}}(y)=\frac{{d\sigma}_{p-p}}{{dy}}(y').
\end{equation}
Here,
\begin{eqnarray}
\frac{{d\sigma}_{p-p}}{{dy}}(y')&=&\rho_{J/\psi}\int^{2m_{D}}_{2m_{c}}dm\frac{2m}{s}
\times[f_{g}(x'_{1g},m^{2})f'_{g}(x'_{2},m^{2})\sigma_{gg}(m_{c}^{2})\nonumber \\
& & +\sum^{}_{q=u,d,s}\{f_{q}(x'_{1},m^{2})f'_{\bar{q}}(x'_{2},m^{2})
+{f_{\bar{q}}(x'_{1q},m^{2})f'_{q}(x'_{2},m^{2})}
\}\sigma_{q\bar{q}}(m_{c}^{2})],
\end{eqnarray}
with
\begin{eqnarray}
y'=y+ln(\frac{E+\Delta E_{c\bar{c}}}{E}),
\end{eqnarray}
\begin{eqnarray}
x'_{1g}=\frac{m}{\sqrt{s}}e^{y'}+\Delta E_{g}/E_{p}, & & x'_{1q}=\frac{m}{\sqrt{s}}e^{y'}+\Delta E_{q}/E_{p},
\end{eqnarray}
and
\begin{eqnarray}
x'_{2}=\frac{m}{\sqrt{s}}e^{-y'}.
\end{eqnarray}

\section{Results and discussion}
In order to determine the value of incoming gluon energy loss parameter $\alpha$,  we give the phenomenological analysis at the leading order for $J/\psi$ production cross section
ratios $R_{W(Fe)/Be}(x_{F})$:
\begin{equation}
\textcolor[rgb]{1.00,0.00,0.00}{R_{W(Fe)/Be}(x_{F})=\frac{d\sigma_{p-W(Fe)}}{d x_{_{\rm
F}}}/\frac{d\sigma_{p-Be}}{d x_{_{\rm F}}},}
\end{equation}
for the E866 experimental data (49 dots) by using the EPS09 nuclear parton distributions [16] together with the energy loss parameter of incident quark ($\beta=1.21\pm0.09$ GeV/fm) determined from the nuclear Drell-Yan experimental
data[26] and the color octet $c\overline{c}$ energy loss ($\alpha=2.97$ GeV/fm) determined in our previous work[15].
By minimizing $\chi^2$ with the CERN subroutine MINUIT
[28] the value of parameter $\alpha$ are extracted: $\alpha=2.18\pm0.14$ GeV/fm.
One standard deviation of the optimum parameter
corresponds to an increase of $\chi^{2}$ by 1 unit from its minimum
$\chi^{2}_{min}$.  The result indicates that the incoming gluons lose more energy than the incident quarks in $J/\psi$ production from p-A collisions, which is in accord with the prediction that gluons lose more energy than quarks because of the stronger coupling to the medium[18-22].
In addition, due to the effects of the modification of the gluon parton distribution functions on the nucleus leading to an additional $J/\psi$ suppression in p-A collisions, the EPS09 uncertainties can be the main source of uncertainties associated to our results.

To identify the energy loss effect of the incoming gluon and incident quark on the $J/\psi$ suppression, the theoretical results are compared with E866 experimental data\textcolor[rgb]{1.00,0.00,0.00}{[4]} at $\sqrt{s}=38.7$ GeV in figure 1 (figure 2), RHIC experimental data[9] at $\sqrt{s}=200$ GeV in figure 3, and LHC experimental data[7,8] at $\sqrt{s}=5.0$ TeV in figure 4, respectively.
The dotted, dashed and solid lines correspond to the results given without initial state energy loss, by considering the incident quark energy loss effect, and the energy loss of incident quark together with incoming gluon energy loss.

\begin{figure}[t,m,b]
\centering
\includegraphics*[width=14cm, height=12cm]{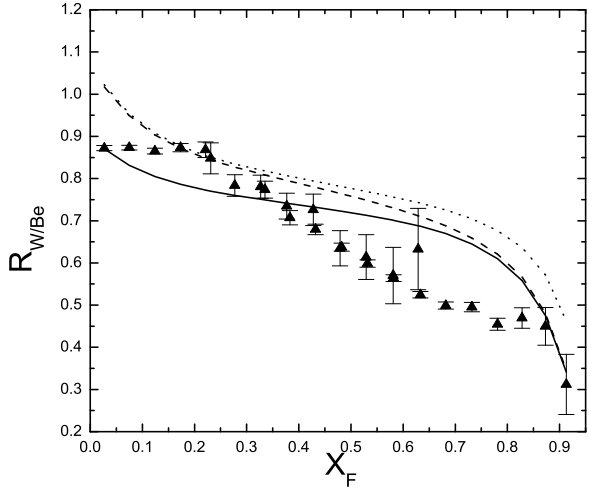}
\vspace{-4.0cm} \caption{The calculated $J/\psi$ production
cross section ratios $R_{W/Be}(x_{F})$ without the initial state energy loss(dotted
line), by considering the incident quark energy loss effect( dashed line), and the energy loss of incoming quark and gluon (solid line). The solid triangles are the
E866 experimental data\textcolor[rgb]{1.00,0.00,0.00}{[4]}.}
\end{figure}
\begin{figure}[t,m,b]
\centering
\includegraphics*[width=14cm, height=12cm]{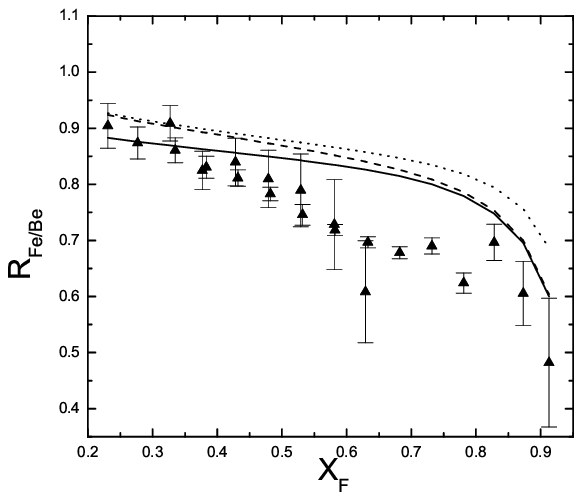}
\vspace{-4.0cm} \caption{The calculated $J/\psi$ production
cross section ratios $R_{Fe/Be}(x_{F})$. The other comments are the same as those in Fig. 1.}
\end{figure}

 As can be seen in figure 1 (figure 2), the nuclear suppression due to the incident quark energy loss can be negligible in the region $x_{F}<0.3$, increases gradually in $x_{F}<0.8$, and becomes steeper in $x_{F}>0.8$.  But on the contrary, the suppression from the energy loss effect of incoming gluon is much steeper in the region $x_{F}<0.3$, reduces gradually in $x_{F}<0.8$, and becomes ignored in $x_{F}>0.8$. It is clear that the incident gluon energy loss plays an important role in the suppression of $J/\psi$ production cross section ratios $R_{W(Fe)/Be}(x_{F})$ in the small $x_{F}$ region (especially for $x_{F}< 0.3$), and the energy loss effect of incoming quark is obvious in the large $x_{F}$ region (especially for $x_{F}>0.8$). We can see that the experimental data on $J/\psi$ production at E866 energy ($\sqrt{s}=38.7$ GeV) can give a best test for the identity of the incident parton which loses its energy in the nuclear medium.

\begin{figure}[t,m,b]
\centering
\includegraphics*[width=14cm, height=12cm]{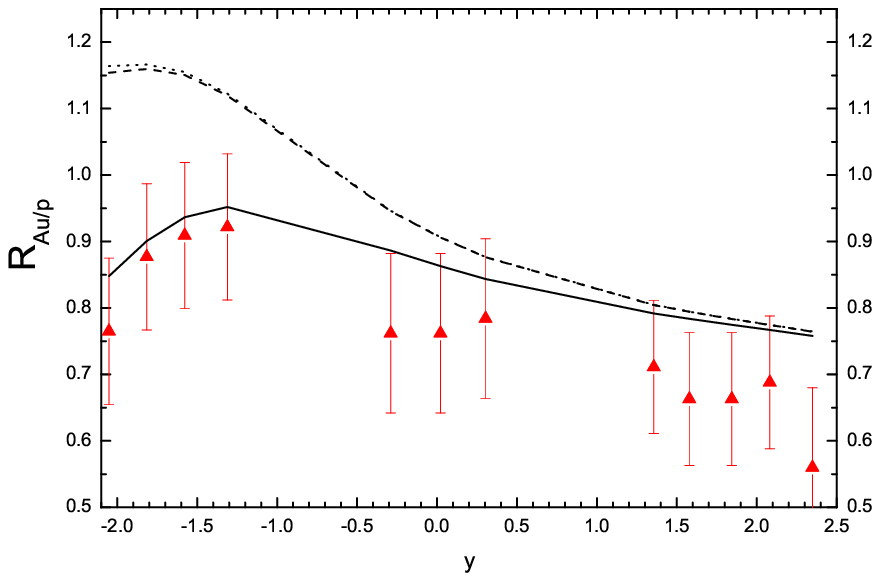}
\vspace{-4.0cm} \caption{The calculated $J/\psi$ production
cross section ratios $R_{Au/p}(y)$ without the initial state energy loss(dotted
line), by considering the incident quark energy loss effect( dashed line), and the energy loss of incoming quark and gluon (solid line). The solid triangles are the
RHIC experimental data[9].}
\end{figure}
\begin{figure}[t,m,b]
\centering
\includegraphics*[width=14cm, height=12cm]{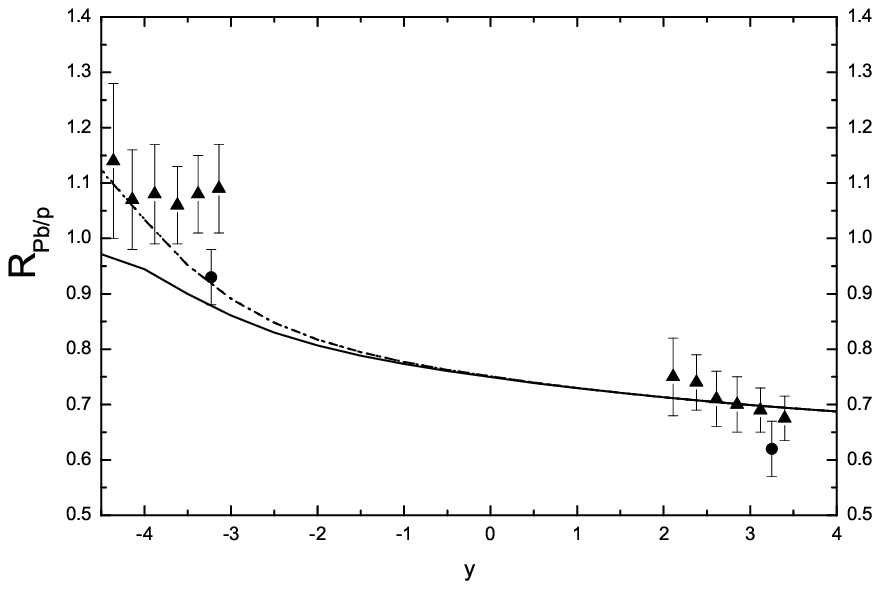}
\vspace{-4.0cm} \caption{The calculated $J/\psi$ production
cross section ratios $R_{Pb/p}(y)$. The solid triangles and filled circles correspond to the experimental data form ALICE Collaboration [7] and LHCb Collaboration [8] at LHC, respectively. The other comments are the same as those in Fig. 1. }
\end{figure}

In figure 3 and figure 4, the theoretical results \textcolor[rgb]{1.00,0.00,0.00}{about $J/\psi$ production
cross section ratios $R_{Au(Pb)/p}$ as a function of $y$} are compared with RHIC[9] and LHC[7,8] experimental data, respectively. From figure 3 we can see that the dotted line and the dashed line appears to overlap, which indicate that the energy loss effect due to the incoming quark plays no role on the $J/\psi$ production at the RHIC energy. In contrast, the nuclear suppression due to the incident gluon energy loss is  obvious especially in the range  $y < -1.5$, reduces gradually with the increase of $y$, and becomes ignored in $y>2.0$. As also can be seen in figure 4, the incident quark energy loss effect has little impact on the $J/\psi$ production cross section
ratio $R_{Pb/p}(y)$ at the LHC energy, and the energy loss due to the incoming gluon plays an important role on the nuclear suppression especially in the range $y < -3.5$ , reduces gradually with the increase of $y$, and becomes ignored in $y>-1.5$. In the present work, it is found that the energy loss of incoming gluon plays an important role on the suppression of $J/\psi$ production in a wide energy range from $\sqrt{s}=38.7$ GeV to $\sqrt{s}=5.0$ TeV, and the influence of incident quark energy loss can be ignored for high energy, such as at RHIC energy and LHC energy.

\section{ Summary }

Following our previous work[14,15], we study the energy loss effect of incoming gluon from $J/\psi$ production in p-A (or d-A) collisions.
By means of the EPS09 nuclear parton
distributions[16] together with the energy loss of the incident quark ($dE/dL = 1.21\pm0.09$ GeV/fm determined in our work [26]) and color octet
$c\overline{c}$ ($dE/dL = 2.97\pm0.74$ GeV/fm determined in our study [15]), we give the phenomenological analysis at the leading order for $J/\psi$ production cross section
ratios $R_{W(Fe)/Be}(x_{F})$ for the E866 experimental data (49 dots) and extract the gluon mean energy loss per
unit path length $dE/dL = 2.18\pm0.14$ GeV/fm by minimizing $\chi^2$ with the CERN subroutine MINUIT
[28].  This result indicates that the incoming gluons lose more energy than the incident quarks, which supports the prediction that gluons lose more energy than quarks because of the stronger coupling to the medium[18-22].
In addition, the EPS09 uncertainties can be the main source of uncertainties associated to our results, owing to the effects of the modification of the gluon parton distribution functions on the nucleus leading to an additional $J/\psi$ suppression in p-A collisions. To identify the energy loss effect of the incoming gluon and quark on the $J/\psi$ suppression, the theoretical results are compared with E866\textcolor[rgb]{1.00,0.00,0.00}{[4]}, RHIC[9], and LHC[7,8] experimental data. We find that the nuclear suppression due to the incident gluon (quark) energy loss reduces (increases) with the increase of the kinematic variable $x_{F}$ (or $y$). The energy loss of incoming gluon plays an important role on the suppression of $J/\psi$ production in a wide energy range from $\sqrt{s}=38.7$ GeV to $\sqrt{s}=5.0$ TeV, and the influence of incident quark energy loss can be ignored for high energy, such as at RHIC energy and LHC energy.


\end{document}